# Braiding rule of boundary Majorana-like zero mode

Qiyun Ma[1], Hailong He[1], Meng Xiao[1, 2,*], Zhengyou Liu[1, 3,*]

[1]Key Laboratory of Artificial Micro- and Nano-structures of Ministry of Education and School of Physics and Technology, Wuhan University, Wuhan 430072, China

[2]Wuhan Institute of Quantum Technology, Wuhan 430206, China

[3]Institute for Advanced Studies, Wuhan University, Wuhan 430072, China

[*]Corresponding author: phmxiao@whu.edu.cn, zyliu@whu.edu.cn.

The study of topological states has become an important topic in both solid-state systems and artificial structures such as photonic crystals and phononic crystals. Among them, Majorana zero modes, which exhibit nontrivial braiding process, have attracted extensive research interest. The analog of Majorana zero modes in classical waves, or the Majorana-like zero modes (MLZMs), have also got a lot of attention recently. However, the vast majority of previous works concerned with MLZMs that were bounded to vortexes inside the bulk. Here in this work, we unveil the braiding rule of MLZMs that are tunable around the boundary of a vortex-texture Kekulé modulated graphene. We show that the existence of these zero-dimensional boundary MLZMs is protected by the Zak phase of 1D boundary states. As such, we are able to construct multiple MLZMs and analyze the corresponding braiding process. In addition, we also provide an implementation scheme of the boundary MLZMs in acoustic crystals. The tunability of the boundary MLZMs proposed herein offer new freedom for topological states braiding in both solid-state systems and artificial structures.

Topological phases of matter have been a flourishing research field in condensed matter physics in recent years [1-3]. Topological phases support topological states that are robust against perturbations that keep the topology unchanged. This characteristic makes topological states highly promising for a wide range of applications. Besides edge and surface modes in first-order topological phases [4-8], and corner and hinge states in higher-order topological phases [9-13], topological states can also be found at dislocations [14-17], disclinations [18-20] and vortex texture modulations [21-22] that behave as real-space topological defects inside a bulk lattice. Among them, Majorana zero modes (MZMs), a peculiar type of topological zero-energy modes, possess unique features [23-26]. For instance, the adiabatic exchanging of two MZMs, termed braiding, leads to certain unitary transformations imposed on their wavefunctions, which implies that the braiding process is of non-Abelian nature, i.e., changing the order of braiding leads to different final states [27,28]. This unique property holds promise for applications in the realm of quantum information storage and topologically robust quantum computation [29]. Several models and systems have been proposed to realize MZMs [30-37]. However, experimental realizations and non-abelian braiding of MZMs in solid-state systems are still in process [38].

In recent years, the topological zero modes possessed by some artificial lattices have been revealed to share similar properties with the MZMs. [39-42] For instance, classical systems whose Hamiltonians have the same form as the Kitaev model can also lead to topological zero modes. [43-45] In 2007, Hou *et al.* [22] proposed that a fractionally charged topological zero mode exists at the vortex center of vortex-texture Kekulé modulated graphene structure, whose wavefunction shares the same form as that of the MZMs bounded at the vortex in *p*-wave superconductors [27]. Soon after, this idea had been generalized to photonic [46-52], phononic [53,54], and other classical systems [55-57], and the topological zero modes and the braiding of pairs of them are demonstrated. [46] In addition, the unique properties of these topological zero modes have been utilized to achieve optical vortex lasing [51], topological photonic crystal fiber [48], etc. [57] These analogues of MZMs are termed as Majorana-like zero modes (MLZMs) to distinguish from the MZMs in solid state systems. However, all these MLZMs are bounded at vortex centers, and their locations are hardly tunable hence limiting the applications of MLZMs. There have been very primitive conjectures that there is another complementary MLZM for each MLZM at the vortex center based on vortex charge conservation argument. [22,46] However, the detailed mechanism, wave functions, and braiding rule of the complementary MLZMs have not been studied.

In this work, we investigate the MLZMs that are localized at the boundary (denoted as BMLZM for simplicity) of the vortex-texture Kekulé modulated graphene. We point out that there should always be another accompanying BMLZM associated with the MLZM at the vortex center (CMLZM). While keeping the location of the CMLZM unchanged, the location of the BMLZM is tunable by changing the initial offset of the Kekulé modulation phase. To unveil the underlying mechanism of the BMLZM, we study the domain wall states between two domains with uniform-Kekulé-texture modulated graphene. We show that the BMLZM is a zero-dimensional domain wall state that is localized along both the domain wall and boundary. The emergence of the BMLZMs can be traced back to the change of Zak phases of the boundary modes at a semi-infinite bulk as a function of the initial offset phase. Utilizing the reconfiguration of the Kekulé modulation phase, we propose several schemes to braid multiple MLZMs. We also propose a scheme for realizing the MLZMs in an acoustic system.

We start with a tight-binding model [Fig. 1(a)] with a graphene lattice modulated by a vortex-texture Kekulé distortion, whose Hamiltonian reads [22]

$$H = \sum_{r\in\Lambda_a}\sum_{i=1}^{3}(t + \delta t_{r,i})a_r^\dagger b_{r+s_i} + \text{H.c.}, \quad (1)$$

where $a_r$ ($b_{r+s_i}$) is the annihilation operator at site $r$ ($r + s_i$) belonging to sublattice $\Lambda_a$ ($\Lambda_b$). The vector $s_i$ with $i = 1,2,3$ connects points at site $r$ to the three nearest-neighbor lattice sites. The original graphene hopping strength $t$ is modulated by an additional hopping term $\delta t_{r,i} = \Delta_0 e^{i(K_+ \cdot s_i + (K_+ - K_-)\cdot r + \phi_r)}/3 + \text{c.c.}$, where $K_\pm(0, \pm\frac{4\pi}{3a})$ labels the original Dirac points of the unmodulated graphene lattice at the boundary of the first Brillouin zone, and the modulation phase $\phi_r = \theta_r + \alpha_0$ contains the contribution of the azimuth angle $\theta_r$ of site $r$ relative to the vortex center and an initial offset $\alpha_0$. The lattice is terminated at three armchair boundaries whose outer normal directions are all parallel to $\Gamma K_+$, as depicted in Fig. 1(a). Unlike Zigzag boundary and other irregular boundaries, the armchair boundary considered herein is free of trivial zero energy modes. For reference, the situations for systems terminated with other types of boundaries are discussed in the Supplemental Materials (S. M.) Sec. I [58]. The vortex center is chosen at the center of this triangular shape, and the inset in Fig. 1(a) shows the distribution of $\phi_r$ for $\alpha_0 = 0$. The number of lattices used in the calculation is not shown explicitly in Fig. 1(a). We use $N_b$ to denote the number of lattices along each armchair boundary, e.g., $N_b = 4$ in Fig. 1(a). The eigenvalue spectrum of a vortex texture Kekulé modulated graphene with $\alpha_0 = 0$ and $N_b = 18$ is shown in Fig. 1(b). There are two zero energy modes in the band gap. One is bounded at the center of the Kekulé vortex, whose eigenfield is only located on sublattice $\Lambda_b$ [Fig. 1(c)], while the other one (i.e., BMLZM) is bounded at the boundary and is located only on sublattice $\Lambda_a$ [Fig. 1(d)]. The insets in Fig. 1 sketch the positions of the MLZMs.

The location of the CMLZM is independent of $\alpha_0$ while that of the BMLZM depends on $\alpha_0$. To demonstrate this property, we adiabatically increase the value of $\alpha_0$ from 0 to $2\pi$ and plot a few typical eigenfield distributions in Figs. 1(d)-(f) (More details are in the S. M. Sec. II [58]). The BMLZM rotates clockwise along the boundary and returns to the initial place as $\alpha_0$ increased to $2\pi$ [see Figs. 1(d) and 1(f)]. An important observation of the BMLZM is that its location always centers at position with $\phi_r = 0.873\pi$ regardless of the value of initial offset $\alpha_0$. Notably, the BMLZM will acquire an additional $\pi$ phase when it crosses the corner of the lattice, as shown in Fig. 1(e). Specifically, the field components of the sites closest to the corner have different signs on the different sides of the corner. This sign change at the corner is due to the fact that the wavefunction should satisfy the continuity boundary condition at both sides of the lattice corner (S. M. Sec. III). Thus, the BMLZM acquires an additional $3\pi$ phase at the end of an adiabatic evolution with $\alpha_0 = 2\pi$. Considering a $2\pi$ redundant phase freedom, the BMLZM acquires a $\pi$ phase shift, similar to the braiding process of two CMLZMs as discussed in previous works [46, 50].

For a large enough system, the modulation phase at the boundary can be effectively regarded as locally uniform. To simplify our analysis, we start with the case with uniform-texture Kekulé modulation phase $\phi$ which only depends on the initial offset $\alpha_0$, i.e., the $\theta_r$ term is set as zero in $\phi_r$. As shown in Fig. 2(a), we consider a ribbon lattice which is periodic in the *x* direction and large enough in the *y* direction. The outer normal of the upper armchair edge boundary is parallel to $\Gamma K_+$ (the same as the boundaries in Fig. 1), while that of the lower armchair edge boundary is parallel to $-\Gamma K_+$. A more detailed discussion about the two boundaries can be found in S. M. Sec. IV [58]. Applying the Bloch period condition in the $x$ direction, we calculate the energy bands of the ribbon lattice for $\phi$ with $k_x = 0$, as shown in Fig. 2(b). The uniform-texture Kekulé modulation couples the two Dirac points at K and K' points, and a bulk gap whose size depends on the modulation strength opens [22,59,60]. Besides the bulk states (yellow shaded region), there are two boundary modes (dark green and brown bold lines) on each armchair boundary, and they form one-dimensional Dirac points at $\phi = \pm 0.873\pi$ for the upper and lower boundary states, respectively. The boundary states for $\phi \neq \pm 0.873\pi$ always remain gapped for all $k_x$. We will focus on the upper boundary hereafter.

Figures 2(c) and 2(d) show the energy band of the ribbon lattice for $k_x$ with $\phi = 0.973\pi$ and $\phi = 0.773\pi$, respectively. The locations of these two $\phi$ values are indicated by the two vertical dashed lines in Fig. 2(b), which are purposely chosen to be on different sides of the Dirac point of the upper boundary states. The Zak phases of the boundary bands are also shown in Figs. 2(c) and 2(d), and clearly a (0, 0) to ($\pi$, $\pi$) topological phase transition occurs. Such a phase transition is similar to the well-known Jackiw-Rebbi mechanism [61]. Accordingly, a boundary domain wall state emerges at the domain wall of two regions with $\phi$ at two sides of the boundary Dirac point. To demonstrate this point, we calculate the band structure with a domain wall formed by uniform-texture Kekulé modulation phase $\phi = 0.973\pi$ and $\phi = 0.773\pi$, as shown in Fig. 2(e). Periodic boundary condition is still kept in the *x* direction, and the number of lattices in the *y* direction is finite while large enough. The energy spectrum of the lattice in Fig. 2(e) is shown in Fig. 2(f). Two zero energy modes emerge in the band gap, corresponding to the boundary domain wall states. Note that there is no topology phase transition at the lower boundary, and hence these two zero energy modes are all localized at the upper boundary. One is localized at the middle domain wall and the other at the outer domain wall since a periodic boundary condition is applied. Figure 2(e) shows the eigenfield of the zero-energy mode localized at the middle domain wall, which exhibits almost the same field distribution as the BMLZM in Fig. 1. Hence, the discussions above explain the topological origin and as well as the localization of the BMLZM at $\phi_r = 0.873\pi$.

The similarity of the zero-energy mode of the Kekulé modulated graphene lattice and the Majorana zero-energy modes can be revealed with the effective Hamiltonian describing the low energy limit. The effective $\boldsymbol{k} \cdot \boldsymbol{p}$ Hamiltonian containing only the first-order terms near $\Gamma$ point is

$$H = [u_b^\dagger, u_a^\dagger, v_a^\dagger, v_b^\dagger] \frac{3t}{2} \begin{pmatrix} & -iAt\partial_{\bar{z}} & \Delta & \\ -iAt\partial_z & & & \Delta \\ \bar{\Delta} & & & iAt\partial_{\bar{z}} \\ & \bar{\Delta} & iAt\partial_z & \end{pmatrix} \begin{bmatrix} u_b \\ u_a \\ v_a \\ v_b \end{bmatrix}, \quad (2)$$

where $\Delta = \Delta_0 e^{i\phi_r}$ is the Kekulé modulation with phase $\alpha_0$, $\partial_z = i\partial_x - \partial_y$, the overline denotes complex conjugate, $A = \sqrt{3}a$ is the lattice constant, and $u_a$, $u_b$ ($v_a$, $v_b$) are the annihilation operators at Dirac point $K_+$ ($K_-$) for the corresponding sublattice (More details are in S. M. Sec. V [58]). Notably, this Hamiltonian has the similar form with the BdG Hamiltonian in topological superconductor for the state at the vortex center [26,27], demonstrating the connection between the CMLZM and MZM. In fact, the parametric dependence on $\alpha_0$ of the CMLZM here is identical to the MZM on the phase of the superconducting order parameter. Considering the BMLZM, the solution of the effective Hamiltonian in Eq. (2) with zero energy for the upper boundary has the form

$$u_a = e^{\frac{\Delta_0 \cos(\phi_r - \phi')}{At}(x-x_0)} e^{-\frac{\Delta_0 \sin(\phi_r - \phi')}{At}(y-y_0)},$$

$$u_b = e^{-\frac{\Delta_0 \cos(\phi_r - \phi')}{At}(x-x_0)} e^{-\frac{\Delta_0 \sin(\phi_r - \phi')}{At}(y-y_0)}, \quad (3)$$

$$v_a = \pm u_a^*, \text{ and } v_b = \pm u_b^*,$$

where $\phi'$ is a parameter determined by the boundary condition. A comparation between the solutions of Eq. (3) and the tight-binding model is shown in S. M. Sec. VI [58]. These solutions give rise to two zero state defined by $\psi_A = u_a + v_a$ and $\psi_B = u_b + v_b$, which are supported on the corresponding sublattices, respectively. However, only one solution can be normalized depending on specific boundary condition at the domain wall (More details are in S. M. Sec. V [58]). Note here, except for the specific form of the wave function, the intrinsic symmetry possessed by Eq. (3) is the same as the CMLZMs studied in previous works. [22,47,53] In S. M. Sec. VII, we show that a CMLZM can be annihilated with another BMLZM with the opposite topological charge, proving that CMLZMs and BMLZMs share the same topological origin.

Multiple MLZMs can be harnessed to perform interesting braiding processes. The braiding among CMLZMs has been investigated in Refs. [46,50]. Figure 1 shows the braiding between a CMLZM and a BMLZM, where the adiabatic change of modulation phase $\phi_r$ from 0 to $2\pi$ introduces a $\pi$ phase on both the CMLZM and the BMLZM. We now proceed to demonstrate the braiding among BMLZMs. The generation of multiple BMLZMs can be easily achieved by adjusting the number of domain walls through the distribution of $\phi_r$ in the lattice. Note here, since all the BMLZMs are supported on the same sublattice and our system exhibits chiral symmetry, these BMLZMs do not couple with each other, even when they are located in close proximity. Thus, the energies of these BMLZMs remain zero as long as the distances between BMLZMs and CMLZMs are sufficiently large.

Firstly, we start with the braiding of two BMLZMs locating on the same armchair boundary as illustrated in Fig. 3(a). The corresponding modulation phase distribution is provided in S. M. Sec. VIII. Initially, the two BMLZMs are in the same phase [lower left panel of Fig. 3(a)], denoted as $|\psi_0> = [1,1]^T$. After adiabatically increasing the modulation phase $\phi_r \to \phi_r + 2\pi$, the two BMLZMs rotate clockwise. The original left BMLZM relocates to the initial position of the right BMLZM without any phase change; simultaneously, the right BMLZM goes around the sample, returning to the original position of the left BMLZM with an additional $\pi$ phase. The resulting states of the two BMLZMs after the braiding is thus denoted as $|\psi_T> = [-1,1]^T$, and this braiding process is given by $|\psi_T> = U_2|\psi_0>$ with $U_2 = [0,1;-1,0]$. Hence, as sketched in Fig. 3(a), both BMLZMs acquire an additional $\pi$ phase compared to the initial state, i.e., $|\psi_{2T}> = U_2|\psi_T> = [-1,-1]^T$ after another period of braiding.

Since the BMLZM acquires a $\pi$ phase each time it traverses across a corner, the braiding rule of $n$ BMLZMs here can captured by the braiding matrix $U_n = \sum_{i,j}[U_n]_{i,j}$, where $[U_n]_{1,n} = (-1)^{n_1}$, $[U_n]_{i,i-1} = (-1)^{n_i}$ for $i > 1$ and $[U_n]_{i,j} = 0$ otherwise. Here $n_i$ records the number of times the $i$-th BMLZM crosses the corners during the adiabatic process defined by $\phi_r \to \phi_r + 2\pi$. Since the BMLZMs return to its original configuration after braiding, $\sum_i n_i = 0$ (mod 3). With $U_n$ obtained, now we are able to interpret a few other typical braiding configurations as sketched in Fig. 3(b). (The specific field distributions of these configurations are provided in S. M. Sec. VIII.) Note here, the periods of braiding are different in the middle and lower panels of Fig. 3(b) even though they have the same number of BMLZMs. Hence, different from the braiding rule of CMLZM discussed in previous work, here Fig. 3 shows that the end state after braiding and the period of braiding depends on the specific number of BMLZMs on each armchair boundary.

In conclusion, we have studied in detail the braiding rule of MLZMs of the vortex-texture Kekulé modulated graphene involving both the CMLZMs and BMLZMs. We have proven that the BMLZM is a domain wall state associated with the band inversion of the boundary modes. The tunability and reconfigurability of the BMLZMs are demonstrated by designing the distribution of the Kekulé modulation phase. Multiple nontrivial braiding scenarios with different matrices and braiding periods have been demonstrated. Our system can be implemented in various platforms such as photonic crystals [46-52], phononic crystals [53,54], and circuits [41] In S. M. Sec. IX, we design an acoustic crystal which is able to demonstrate the above braiding rules. Our findings provide new possibilities for manipulating topological zero modes, offering promising candidates for implementing non-abelian braiding within classical waves.

**Acknowledgments**

This work is supported by the National Key Research and Development Program of China (Grants No. 2022YFA1404900, No. 2023YFB2804701), the National Natural Science Foundation of China (Grant No. 12274330)

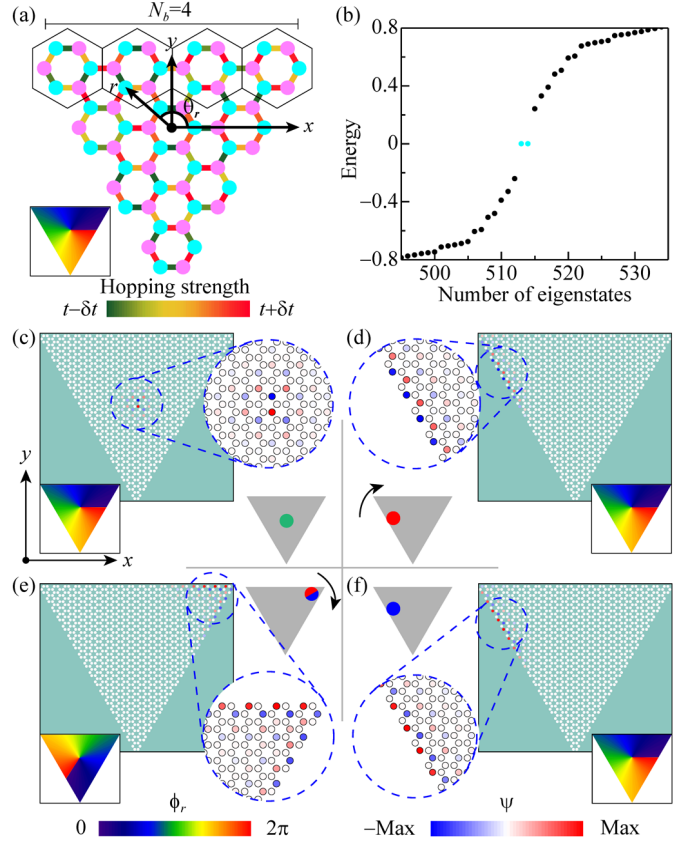

FIG. 1. (a) Tight-binding model of a honeycomb lattice under a Kekulé vortex modulation. The lattice points in cyan (light magenta) belong to sublattice $\Lambda_a$ ($\Lambda_b$). The strength of the hoppings modulated by the Kekulé vortex texture is represented by the color of the bonds. The vortex center is set at the center of the triangular shape whose boundary has $N_b$ lattices. (b) Energy spectrum in the vicinity of zero energy. Cyan points highlight the two MLZMs. (c) Eigenfield distribution of the CMLZM. (d-f) Eigenfield distributions of the BMLZMs for different $\alpha_0$. The insets in the middle sketch the locations of the MLZMs, where the half-red-half-blue disk in (e) highlights the $\pi$ phase jump at the corner. The black arrow marks the evolution direction of the BMLZM with the increasing of $\alpha_0$. Insets of (a), (c)-(f) show the distribution of modulation phase $\phi_r$. In (b-f), $N_b = 18$, $t = 1$ and $\Delta_0 = -1.5$. $\alpha_0 = 0$ in (b-d), $\alpha_0 = 0.7\pi$ in (e) and $\alpha_0 = 2\pi$ in (f).

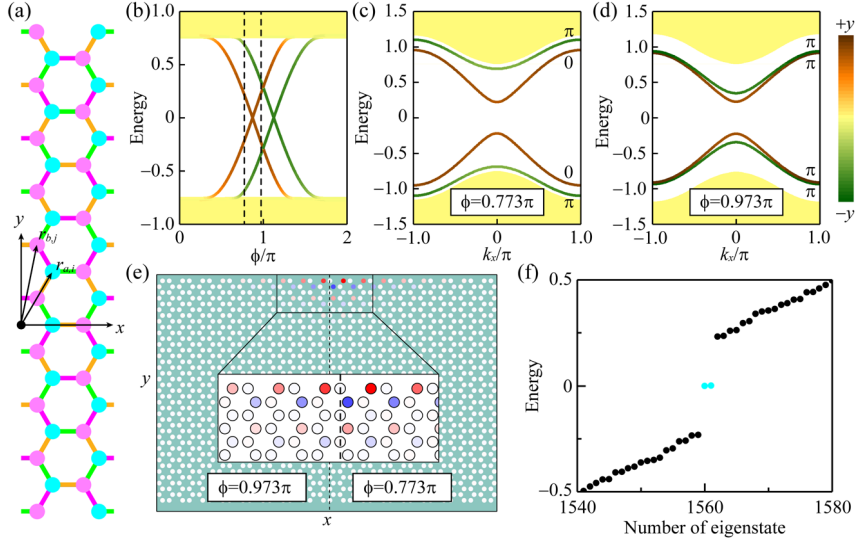

FIG. 2. (a) A ribbon of the lattice with uniform Kekulé modulation denoted by $\phi$. (b) Energy bands of the ribbon structure at $k_x = 0$ as a function of $\phi$. Energy bands of the ribbon structure for $k_x$ at $\phi = 0.773\pi$ (c) and $\phi = 0.973\pi$ (d), where the Zak phases of the boundary modes are also given. In (b-d), the color code represents the location of the eigenstates, brown (dark green) indicates states at the upper (lower) boundary and yellow denotes bulk states. (e) A domain wall formed between two regions with different uniform Kekulé modulation phases, with the corresponding energy spectrum shown in (f). The cyan points represent the two domain-wall zero modes and the field distribution of one is shown in (e). Except for the modulation phase, the parameters used are the same as in Fig. 1. Periodic boundary conditions are applied in the $x$ direction for both (a) and (e).

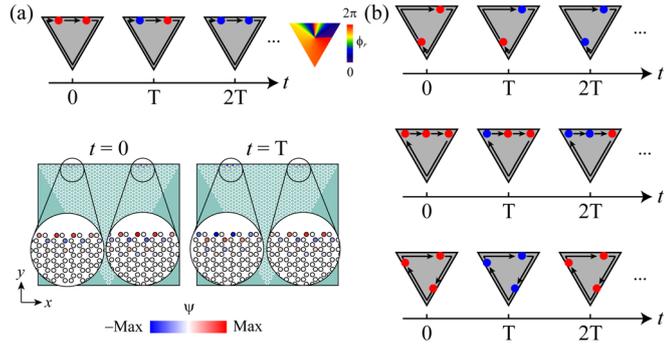

FIG. 3. (a) illustrate the case when two BMLZMs are located at the same boundaries. Here $T$ represents the period for the BMLZMs to rotate back to the initial configuration when the modulation phase $\phi_r \to \phi_r + 2\pi$. The lower panels show the field distributions at $t = 0$ and $t = T$. The red and blue dots in the upper panel sketch the locations of the BMLZMs with the color indicating whether there is a $\pi$ phase change during this braiding process. If the color of a dot changes, then there is $\pi$ phase obtained; otherwise, there is no additional phase. (b) shows another three different braiding configurations. The parameters used and detailed field distributions are provided in S. M. Sec. VIII.